\documentclass[%
 reprint,
 amsmath,amssymb,
 aps,
]{revtex4-1}

\usepackage{graphicx}
\usepackage{dcolumn}
\usepackage{bm}

\begin{document}

\preprint{APS/123-QED}

\title{Glassy dynamics in disordered oscillator chains}

\author{Alen Senanian}
\email{asenania@ucsc.edu}
\author{Onuttom Narayan}%
\email{onarayan@ucsc.edu}
\affiliation{%
Physics Department, University of California, Santa Cruz CA 95064
}%

\date{\today}

\begin{abstract}
The  escape of energy injected into one site in a disordered chain
of nonlinear oscillators is examined numerically. When the disorder
has a `fractal' pattern, the decay of the residual energy at the 
injection site can be fit to a stretched exponential 
with an exponent
that varies continuously with the control parameter. At low
temperature, we see evidence that energy can be trapped for an
infinite time at the original site, i.e. classical many body localization.
\end{abstract}

\maketitle

The Fermi Pasta Ulam (FPU) model of coupled nonlinear
oscillators~\cite{fermi} has served as a testing ground for basis
ideas in statistical mechanics for more than half a century.  After
the initial result that energy did not distribute itself efficiently
between the oscillator modes, there was a huge body of work~\cite{ford},
including the development of related integrable models~\cite{soliton,toda}.
The fact that the FPU model is approximately integrable is generally
believed to result in very long equilibration times, often beyond
what is observable numerically~\cite{bennetin}.  At high energies,
equilibration proceeds efficiently and such metastable behavior is
not seen~\cite{hightemp}. The FPU model has also been used to study
heat conduction in low dimensional systems, where a heat conductivity
that diverges in the thermodynamic limit is seen for various
models~\cite{dhar}.

Even when the FPU model does not equilibrate efficiently, the normal
modes in which the the energy is concentrated are delocalized. It
is possible to construct disordered and linearized versions of the
FPU model for which the normal modes are localized, but the normal
modes are all decoupled from each other. Following the great progress
in the field of many body localization for quantum statistical
systems~\cite{huse}, it is natural to ask if localization of energy
can be achieved in interacting classical systems too. The study of
heat conduction in disordered nonlinear oscillator chains suggests
otherwise~\cite{DL}: even a small amount of nonlinearity is found
numerically to result in normal heat conductivity for large chains.
Although this implies that a localized energy packet will spread
out, it is still possible that a fraction of the energy packet will
remain at its original location.

In this paper, we consider a one-dimensional ring of linear
oscillators, each with a different frequency, in which nearest
neighbors are coupled together with a nonlinear potential.  The
system is initialized by equilibrating at a temperature $T.$
Thereafter, a packet of energy is deposited at one site, and the
system is evolved using molecular dynamics. The excess energy at
the site is measured as a function of time $t.$ If the system
thermalizes, the difference of excess energy $\Delta E(t)$ should
vanish as $t\rightarrow\infty.$

In order to avoid accidental resonances between oscillators that
are far away from each other, or rare regions where clusters of nearby
oscillators are resonant resulting in chaotic spots in the 
dynamics~\cite{basko}, the frequencies $\{\omega_i\}$
have to be chosen judiciously. We choose
the frequencies in a `fractal' manner: a large
gap between the frequencies of sites near each other, and a
small gap between the frequencies of sites far away from
each other, with the gap size decaying as
a power of the distance between the sites.  This is made quantitative
later in this paper.

As the nearest neighbor coupling constant $J$ is lowered at low
temperature, we find that the decay of the excess energy $\Delta
E_J(t)$ slows down substantially. The curves for $\Delta E(t)$ for
various values of $J$ can be superimposed on top of each other if
$\ln t$ and $\Delta E$ are scaled for each $J.$ Thus $\Delta E(t)
= A(J) \hat{\Delta E}(t^{\beta(J)}).$ The numerical results are
consistent with a stretched exponential form for the function
$\hat{\Delta E};$ stretched exponential dynamics are often seen 
in experiments on glassy systems~\cite{angell}, and there are various
theoretical models with traps and a range of time scales that obtain
similar behavior~\cite{phillips}. (A power law form with a cutoff 
can also be fit to the data.)

The smallest $J$ value shows an essentially flat $\Delta E(t),$ and
so we vary the temperature at this $J$ and measure the decay of the
excess energy. The behavior of $\Delta E(t)$ is found to be a
non-monotonic function of $T$: it decays slowly at both high
temperature and low temperature, with a more rapid decay at
intermediate temperatures. In the low temperature regime, the
numerical results indicate that the system freezes at a non-zero
temperature $T_f,$ with $\Delta E_{T < T_f} (t\rightarrow\infty)
\neq 0.$ To our knowledge, this is the first evidence of classical
many body localization.

The Hamiltonian of the chain with `fractal' disorder is
\begin{equation} 
H = \frac{m}{2}\sum_{l=1}^N [\dot x_l^2 + \omega_l^2
x_l^2] - J \sum_{l=1}^N \cos(x_l - x_{l-1}) 
\end{equation} 
with periodic boundary conditions. The particles in the
chain all have equal masses and are tethered to their equilibrium
positions by harmonic springs. The tethering ensures that momentum
conservation is destroyed and there is no
anomalous transport~\cite{campbell,ramaswamy}. The
frequency of each of the tethering harmonic oscillators is different.
When $J = 0,$ the energy is obviously localized at each lattice
site. For $J\neq 0,$ the oscillators are coupled, but if $J$ is
small, one might try to use perturbation theory. If we define
\begin{equation}
a_l(t) = \dot x_l(t) - i \omega_l x_l(t)
\end{equation}
then the dynamical equations can be expressed as 
\begin{equation}
\dot a_l = - i \omega_l a_l + \frac{J}{m} [\sin (x_{l+1} - x_l) + \sin(x_{l-1} - x_l)] 
\label{adyn}
\end{equation}
with the supplementary equation
\begin{equation}
x_l = \frac{1}{2 i \omega_l} [a_l^*(t) - a_l(t)].
\end{equation}
This is now a set of coupled first order differential equations.
The solution to zeroeth order in $J$ is trivial.

To first order in $J,$ $a_l$ is forced by terms that are of the
form $\sim \exp[i (m \omega_{l\pm 1} + n \omega_l) t],$ where $m$
and $n$ are integers. The terms with $m=0,n= -1$ shift the natural
frequency of the oscillator $\omega_l$ by an amount that is $O(J).$
Other terms yield $O(J)$ corrections to $a_l,$ of the form
\begin{equation}
\sim  \frac{J}{ m \omega_{l\pm 1} + (n -1) \omega_l} \exp[i (m \omega_{l\pm 1} + n \omega_l) t].
\label{pert}
\end{equation}
To next order in $J,$ each site is influenced by its next nearest
neighbors, and so on. If there is a near degeneracy between the
frequencies of two sites $l$ and $l+r$ that are $r$ steps apart,
for small amplitudes (i.e. low temperatures), the leading
correction to $a_l$ is of the order of
\begin{equation}
\frac{J^r}{(\omega_l - \omega_{l+r})(\omega_{l+1} - \omega_{l+r})\ldots(\omega_{l+r - 1} - \omega_{l+r})}.
\label{perturb}
\end{equation}
If the $\{\omega_i\}$'s are
chosen randomly, accidental near degeneracies can 
result in a small
denominator in (and a breakdown of)  the perturbation
expansion.
To avoid this, 
we choose the frequencies in the following manner.
Let $N = 2^n.$ First, the
frequencies at all the odd and even sites are set to 1 and 3 respectively. 
Next, the frequencies of successive
pairs of sites are increased or decreased by $\lambda,$
so that the frequencies are $\{1 + \lambda, 3 + \lambda, 1 - \lambda,
3 - \lambda, \ldots\}.$ At the next step, the frequencies of
successive groups of 4 sites are increased or decreased by $\lambda^2.$
This is carried out $n$ times, when all the
$\{\omega_i\}$'s are non-degenerate. The frequency
gap between two sites that are an odd multiple of $2^k$ apart satisfies
\begin{equation}
\delta\omega_{l, l + 2^k} \leq 2 \lambda^k[1 - \lambda - \lambda^2\ldots] 
   =  2 \lambda^k \frac{1 - 2\lambda}{1 - \lambda} = 2\lambda^{k+1}
\end{equation}
for $N\rightarrow\infty$ if $\lambda = (5 -
\sqrt{17})/2.$ Then it is easy to see that if $J << 2 \lambda^2,$
the expression in Eq.(\ref{perturb}) is small for all $r.$

Although this is necessary for the perturbation expansion to be
well behaved, it is not sufficient. Regardless of the relationship
between two frequencies, it is always possible to find values of
$m$ and $n$ in Eq.(\ref{pert}) for which $m \omega^\prime + (n -
1) \omega$ is as small as one wishes. This is the problem of small
denominators that makes the KAM theorem~\cite{KAM} difficult.
However, these terms are higher order in the amplitudes of the
oscillators. One might expect them to be important when the amplitude
of an oscillator happens to be large, but in that case, because the
sine function is bounded, the coupling term on the right hand side
of Eq.(\ref{adyn}) is {\it weak} compared to the $-i\omega_l a_l$
term. This is not a proof that these terms are unimportant, and
therefore we turn to numerical simulations.

Because energy transport in this system is at best very slow, one
has to be careful to bring it to thermal equilibrium before a packet
of energy is injected; simply running molecular dynamics for a long
time is not sufficient. We initialized the velocities from
a Gaussian distribution, and equilibrated the coordinates using
Monte Carlo dynamics (with acceptance rate $\approx 0.5$).
Equilibrium was considered to be achieved when the
virial theorem was satisfied to within 10\%.

Once in equilibrium, a heat packet of magnitude $E_+$ was injected
in the system in the form of equal and opposite momentum between
two neighboring sites. The system was then evolved dynamically with
the Forrest-Ruth algorithm, using a time step of $h = 0.01.$ If we
denote $E_l(t ; k)$ as the energy of sites $l$ and $l + 1$ at time
$t$ when the heat packet was injected at sites $k$ and $k + 1$ at
$t=0$, the residual energy at $l$ is defined as
\begin{equation}
\Delta E_l (t) = {1\over 2}[ E_l(t; l) - E_l(t; l + N/2)]
\end{equation}
so that, if the system were to equilibrate, $\Delta E_l(\infty)$
would be zero.  For the measurements reported here, the mass of
each site was $m = 1,$ the size of the ring was $N=64$ or 128,
and the extra energy injected was  $\Delta E(t=0)  = 5.0$.

\begin{figure}
\includegraphics[scale=0.66]{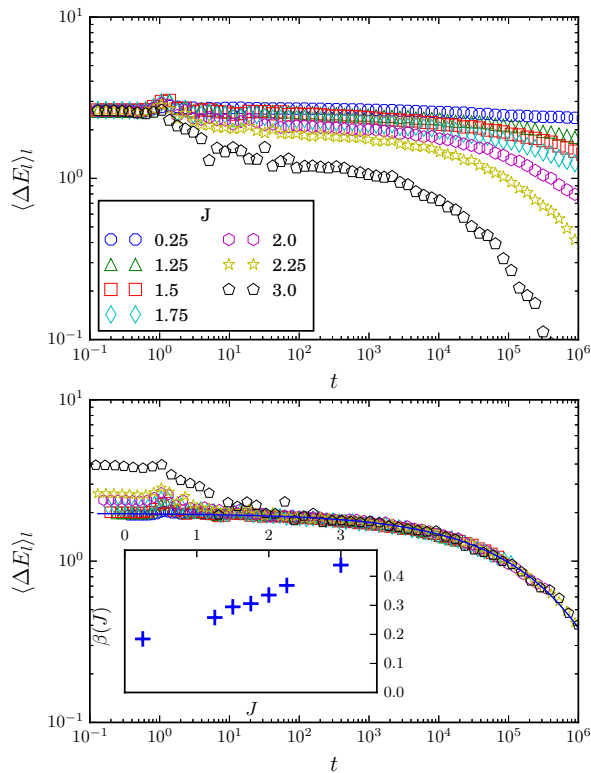}
\caption{\label{fig:fixedT_cos} (Top) Residual
energy as a function of time for the fractal oscillator model with
cosine interactions, with various values of $J$ at $T = 0.25$.
The curves from bottom to top are for decreasing $J$ decreasing.
(Bottom) Rescaled version of the same. A stretched exponential 
curve is shown with a black line. (Bottom inset) 
Time-scaling factor $\beta(J)$ vs $J$.  }
\end{figure}

Figure~\ref{fig:fixedT_cos} shows the residual energy $\langle
\Delta E_l \rangle_l$ averaged over $l,$ as a function of time, for
the fractal oscillator model with various values of the coupling
constant $J$. The system consisted of $N = 64$ sites held at
temperature $T = 0.25$. The brackets $\langle ... \rangle$ denote
averages over initial conditions as well as averages over sites
where the energy was injected and measured. The simulations were
averaged over $30 \times 64$ runs, with 30 runs for each site. As
$J$ was reduced, the dynamics become steadily slower and slower.
The curves collapse onto one another if the horizontal axis is
scaled and the vertical axis is shifted for each curve by a different
amount, and the result fits nicely to a stretched exponential. This
implies that all the curves are of the form $\langle \Delta
E_l(t)\rangle = A(J) \exp[- (t/t_0)^{\beta(J)}],$ with $t_0 = 3
\times 10^5.$ However, the curve for $J=0.25$ is essentially flat,
making it very difficult to determine whether it fits the same
stretched exponential form or if the decay takes infinitely long:
$\Delta E(t\rightarrow\infty)\neq 0.$ In order to elucidate this
further, we hold the coupling constant fixed at $J=0.25$ and vary
the temperature.

\begin{figure}
\includegraphics[scale=0.66]{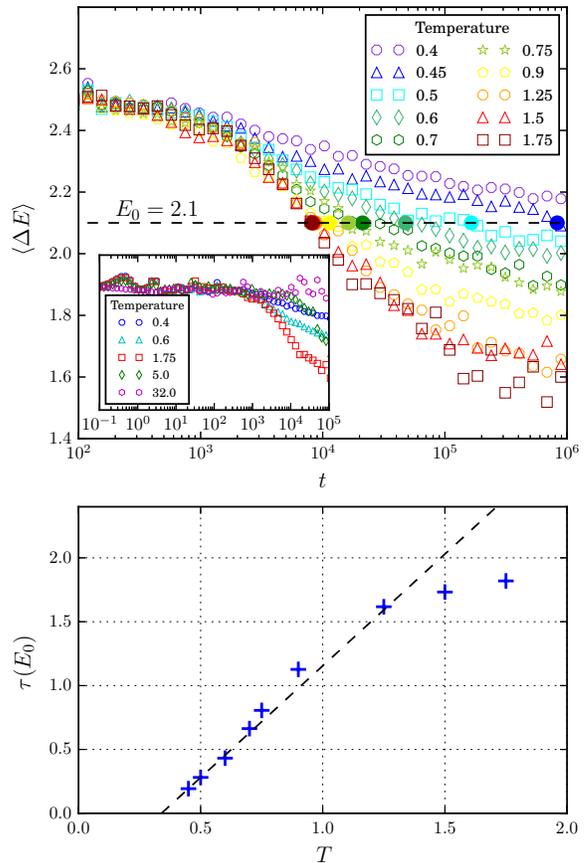}
\caption{\label{fig:varT_cos} (Top inset) Residual energy at sites
$l = 0, 1$ as a function of time for the fractal oscillator model
with cosine interactions, at $J = 0.25$ and
various values of $T.$ (Top) Residual energy for the low temperature
curves from the inset.  Filled circles indicate the points of
intersection $t_I(T)$ of the curves with the dashed line.  The
vertical range of this plot and its inset are identical.  (Bottom)
The parameter $\tau(T) = 1/(\ln t_I(T) - \ln A)$ 
as a function of $T$, with $A$ chosen to yield linear
dependence at low temperature.} 
\end{figure}

Figure \ref{fig:varT_cos} shows the residual energy as a function
of time for the same system (but with $N=128$) at various temperatures
with $J = 0.25.$ Unlike in Figure~\ref{fig:fixedT_cos}, the energy
was only injected at the sites $l=(0,1).$ For $T > 0.50$, the
measurements were averaged over 1200 runs, while only 900 runs were
realized for the lower temperature curves. In addition, the timestep
for $T > 5.0$ were reduced by a factor of 10 to account for the
faster dynamics. Unexpectedly, the decay of the residual energy is
slow at low {\it and\/} high temperatures, but not at intermediate
temperatures.  For the high temperature behavior, we argued earlier
that the coupling between oscillators is weak, and the slow decay
of the residual energy is not surprising.  and at low temperatures,
the linear disordered model (which is localized) has small corrections,
with the same result.

Because of the possibility of metastability in oscillator
chains~\cite{bennetin}, one has to be careful whether the slow decay
really indicates energy being trapped for an infinite time.  Therefore,
the times $t_I(T)$ at which the residual energy drops to $E_0 =
2.1,$ approximately 80\% of the energy originally injected, are
calculated, and found to fit the Vogel Fulcher~\cite{vogel} form
$t_I(T) = A \exp[B/(T - T_f)]$ with $A = 0.47\times 10^5,$ $B=0.57$
and $T_f = 0.34.$ This indicates that a finite residual energy
remains at the original pair of sites when $T < T_f,$ i.e.  $\langle
\Delta E_l(t\rightarrow\infty)\rangle \geq E_0.$

\begin{figure}
\includegraphics[scale=0.66]{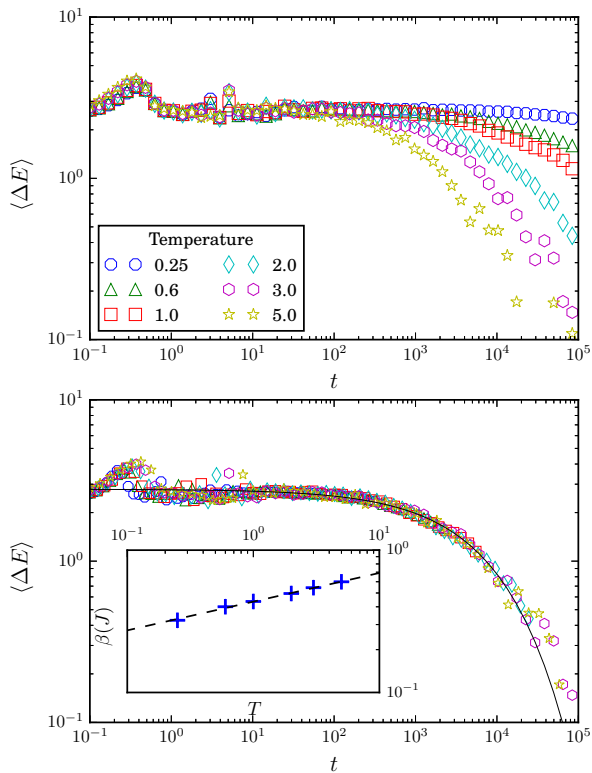}
\caption{\label{fig:varT_x2x4} (Top) $\Delta E(t)$ at sites
$l = 0,1$ as a function of time for the fractal oscillator model
with polynomial interactions, at $J=0.25$
and various values of $T.$ The curves from bottom to top are for
decreasing temperature (Bottom) Rescaled version of the
same data. A stretched exponential is shown 
with a black line.  (Bottom inset) Log-log plot of $\beta(T)$ vs $T.$}
\end{figure}

To confirm our argument that the bounded form of the interaction
potential is (at least in part) responsible for energy localization,
we carried out simulations on the fractal oscillator chain
but with nearest neighbor potential  
$V(u) = J (u^2/2 + u^4/4).$ With $N=64$ and $J=0.25,$ the measurements were
averaged over 1200 runs for $T \leq 1.0$ and 4800 runs for $T >
1.0.$ The results for this system 
are in Figure \ref{fig:varT_x2x4}.  All the curves
collapse on top of each other if the $t$-axis is scaled 
differently for each $T,$ and the result fits to a stretched
exponential (except for large $t$ and $T$). This implies
$\langle\Delta E\rangle \sim A \exp[- (t/t_0)^{\beta(T)}],$
with the best fit values of $A = 2.8$ and $t_0 = 7 \times 10^3.$
The data is consistent with $\beta(T)\sim T^{0.2}.$ Thus
$\beta(T\rightarrow 0)$ appears to be zero, i.e.  one cannot
show that the dynamics freeze at $T\neq 0.$


\begin{figure}[htb]
\includegraphics[scale=0.66]{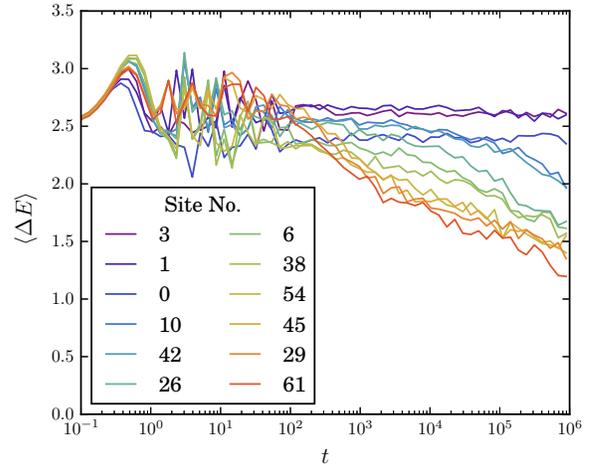}
\caption{Plots of $\Delta E(t)$ when the energy is injected at
various sites in the ring, with $N = 64, T = 0.25$ 
and $J = 1.0.$ 30 runs were averaged for each plot.
The curve labeled $l$ has energy being injected at the
sites $(l, l+1).$ The $l=0$ curve, corresponding to energy 
injection at sites $(0,1)$ which we have studied so far, is third
from the top, i.e. one of the flattest.}
\label{fig:disagg}
\end{figure}

Finally, in Figure~\ref{fig:disagg} we compare $\Delta
E(t)$ when energy is injected at various points in the ring,
showing slow decay for some sites and fast decay for others. 
When energy is
injected into the sites $(0, 1)$, the case discussed so far, the
decay of $\Delta E(t)$ is one of the slowest.

In conclusion, we have studied energy trapping in disordered
classical oscillator chains, with disorder chosen to avoid
resonances. If the coupling between
oscillators has a cosine form, we see evidence that, at low
temperatures, energy can be trapped at a site for infinite 
time, indicating classical many body localization. This is 
not seen when the coupling is polynomial.

We thank Richard Montgomery, David Huse and Sid Nagel for helpful
discussions, and David Huse for suggesting this problem.  O.N. 
thanks the International Center for Theoretical
Sciences (ICTS) for their hospitality during the program Non-equilibrium
Statistical Physics (ICTS/Prog-NESP/2015/10), where some of
this work was carried out.

\end{document}